\documentstyle[12pt,epsfig]{article}

\newlength{\dinwidth}
\newlength{\dinmargin}
\def\lapproxeq{\lower .7ex\hbox{$\;\stackrel{\textstyle
<}{\sim}\;$}}
\def\gapproxeq{\lower .7ex\hbox{$\;\stackrel{\textstyle
>}{\sim}\;$}}

\def\beq{\begin{equation}}
\def\eeq{\end{equation}}
\def\bea{\begin{eqnarray}}
\def\eea{\end{eqnarray}}

\def\GeV{\rm GeV}
\def\msb{\overline{\rm MS}}
\def\dis{{\rm DIS}}

\begin{document}
\titlepage
\begin{flushright}
IPPP/04/64 \\
DCPT/04/128 \\
Cavendish-HEP-2004/27 \\
15th October 2004 \\

\end{flushright}

\vspace*{0.5cm}

\begin{center}
{\Large \bf Physical Gluons and high-$E_T$ Jets. }

\vspace*{1cm}
\textsc{A.D. Martin$^a$, R.G. Roberts$^b$, W.J. Stirling$^a$
and R.S. Thorne$^{c,}$\footnote{Royal Society University Research Fellow.}} \\

\vspace*{0.5cm} $^a$ Institute for Particle Physics Phenomenology,
University of Durham, DH1 3LE, UK \\
$^b$ Rutherford Appleton Laboratory, Chilton, Didcot, Oxon, OX11 0QX, UK \\
$^c$ Cavendish Laboratory, University of Cambridge, \\ Madingley Road,
Cambridge, CB3 0HE, UK
\end{center}

\vspace*{0.5cm}

\begin{abstract}
We propose a more physical parameterization of the gluon distribution for global parton analyses
of deep inelastic and related hard scattering data.  In the new
parameterization the gluon distribution at large $x$ in the $\msb$-scheme 
is driven by the valence quarks, which
naturally produces a shoulder-like form at high $x$, and hence
produces a better description of the Tevatron inclusive jet data.  We perform the new
analysis at both NLO and NNLO.  The improvement is found to be even better at NNLO than at NLO.
We make available the new sets of NLO and NNLO partons, which we denote by MRST2004.
\end{abstract}

%\newpage
\vspace*{0.5cm}
%\section{Introduction}

A detailed knowledge of the partonic structure of the proton is an essential ingredient in
the analysis of hard scattering data from $pp$ or $p\bar{p}$ or $ep$ high energy collisions.
The parton distributions are determined by a global analysis of a wide range of deep inelastic
and related hard scattering data.  The Bjorken $x$ dependence of the distributions is
parameterized at some low scale, and a fixed order (either LO or NLO or NNLO) DGLAP evolution
performed to specify the distributions at the higher scales where data exist.  A global fit
to the data then determines the parameters of the input distributions, see, for example,
Refs.~\cite{MRST2001,CTEQ6}.  The uncertainties in the resulting distributions have
been the subject of much detailed study; see, for
example, Refs.~\cite{CTEQLag, MRSTerror1, MRSTerror2}.  The gluon
distribution at high $x$, $x \gapproxeq 0.3$ is particularly ill-determined.   Indeed,
in the past, this ambiguity has been exploited to describe `anomalous' behaviour of
the inclusive jet distribution observed at high $E_T$ at the Tevatron.

\begin{figure}
\begin{center}
\centerline{\epsfxsize=0.65\textwidth\epsfbox{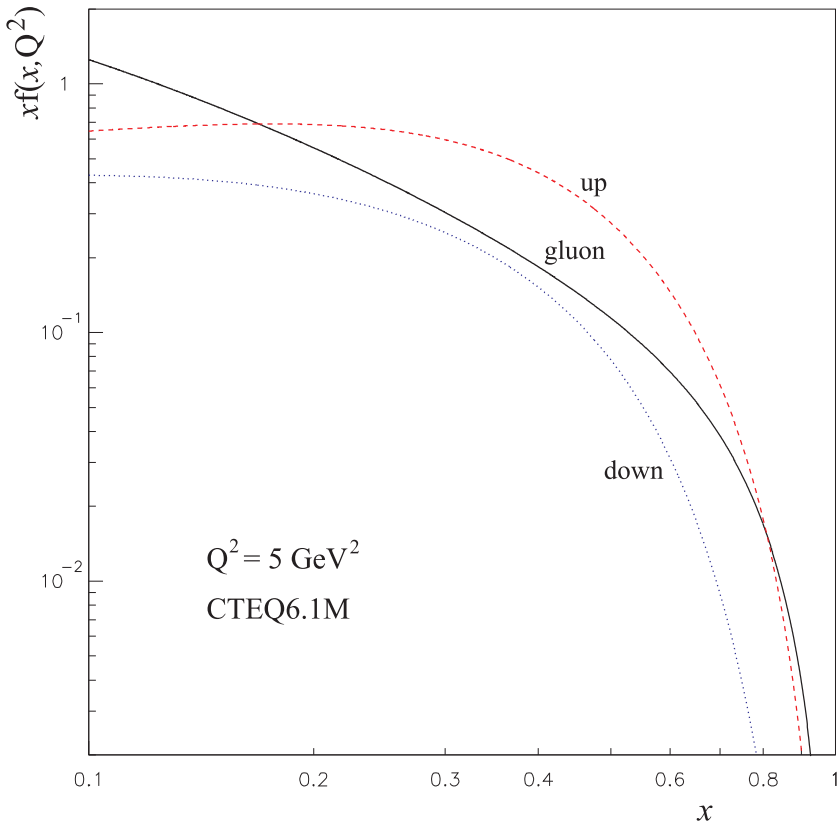}}
\caption{The $x$ behaviour of CTEQ6.1M parton distributions,
$xf(x,Q^2)$, at $Q^2=5~{\rm GeV}^2$.}
\vspace{-1cm}
\label{fig:c6}
\end{center}
\end{figure}

It is informative to illustrate the present situation for high $x$ gluons and the Tevatron jet
data in both the CTEQ and MRST global analyses.   First, we note that the simple spectator counting
rules \cite{BF} predict the following behaviour at high $x$
\beq
\label{eq:count}
q_{\rm val}~\sim~(1-x)^3,~~~~~~~~~~~g(x)~\sim~(1-x)^5,
\eeq
for valence quarks and the gluon respectively.
>From Fig.~\ref{fig:c6} we see\footnote{Such plots can be readily obtained from
http://durpdg.dur.ac.uk/hepdata/pdf3.html}
that this behaviour is not true for CTEQ6.1M (NLO) partons \cite{CTEQ6}.
The gluon is harder than both the up and the down quark distributions as $x \to 1$,
which results in a good fit to the Tevatron jet data.
On the other hand, the MRST parameterizations do not naturally allow
such a hard gluon and, as a consequence the description of
the jet data is not quite so good, the $\chi^2$ being about 30 units higher.  
In fact we have noticed that the problem is worse in the NNLO fit,
than in the NLO analysis.  The NNLO coefficient
functions are positive for $F_2$ at the largest $x$, leading to smaller quarks and a larger gluon
is consequently needed for a good fit.

\begin{figure}
\begin{center}
\centerline{\epsfxsize=0.65\textwidth\epsfbox{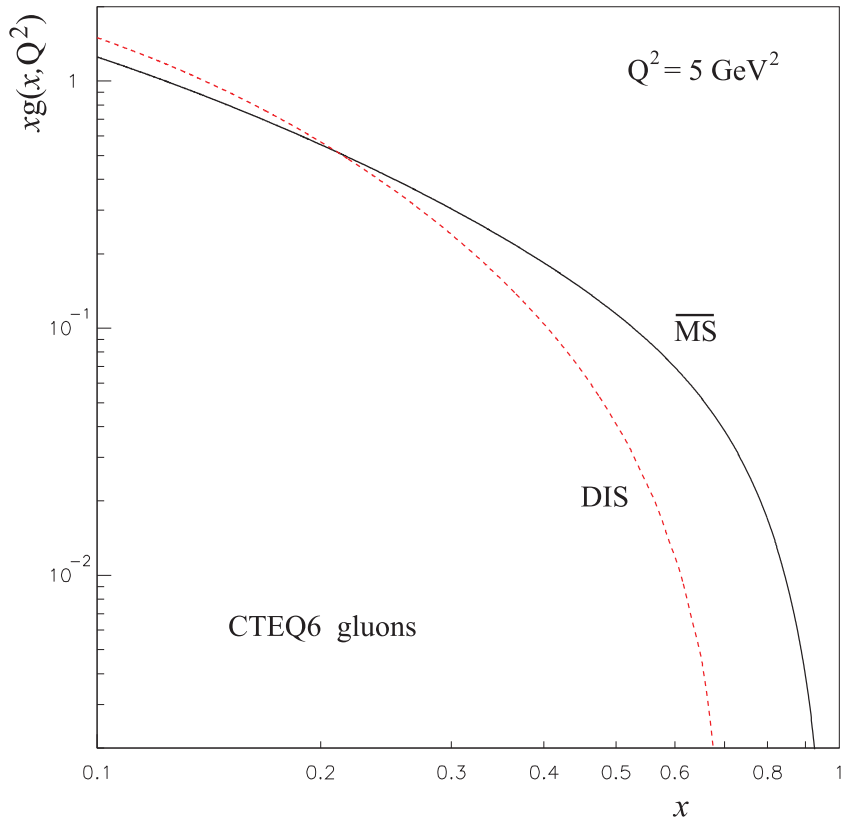}}
\caption{The $x$ behaviour of the CTEQ6.1M and CTEQ6D gluon distributions,
$xg(x,Q^2)$, at $Q^2=5~{\rm GeV}^2$, obtained from global fits using
the $\msb$ and DIS factorization schemes respectively.}
\vspace{-1cm}
\label{fig:c6DM}
\end{center}
\end{figure}

Sometime ago Klasen and Kramer \cite{KK} noticed that the description of the jet data was better
in the DIS factorization scheme than in the $\msb$ scheme. This is for reasons
which we will discuss in a moment.  Note that the latter scheme is the default 
adopted in the global analyses.  Of course, in principle, it should not matter which scheme is
used.  We can readily transform the partons from one scheme to the other without changing
the observables.\footnote{Strictly speaking this is only the case if the NLO, and higher order, 
splitting functions are not exponentiated in the solution to the renormalization group equations.
However, when using the $x$-space evolution programs these terms are exponentiated, so
some higher order terms are introduced. As a consequence a scheme difference
due to these extra terms appears. Nevertheless, this is a small effect, and unrelated
to the results that we highlight in this paper.}
However, in practice, the behaviour of a parton can have a particularly
simple parameterization in one scheme and much more structure in the other scheme.
Since the number of parameters is limited, it is clear
that better fits can occur in the scheme in which the parton has the smoother distribution,
particularly if the structure is difficult to mimic using a particular parameterization.   We shall
see that this applies to the behaviour of the gluon at high $x$.  The first hint that this
might occur can seen from the comparison of the CTEQ6 gluons obtained from separate global
analyses performed first in the $\msb$ scheme and then in the DIS scheme.  Fig.~\ref{fig:c6DM}
shows that the DIS gluon is
far softer than the $\msb$ gluon.  Both are smooth, although a transformation from one to
the other would result in some structure. However, the important point to note is the 
qualitatively completely different behaviour in the two schemes. 

The MRST analyses are performed in the $\msb$ scheme, and the partons are
then transformed to obtain the distributions in the DIS scheme.  It is found that
the gluon becomes negative at high $x$ in the DIS scheme.  All the above observations
indicate that is desirable to look more carefully at the parameterization which
describes the high $x$ behaviour of the gluon.

Indeed, we are led to propose a new approach to the treatment of the gluon distribution at
high $x$.  First we note the general form of the transformation which expresses the
partons in the DIS factorization scheme in terms of those in the $\msb$ scheme \cite{DISdef}.
Schematically we have
\beq
q^{\rm DIS}~~=~~q^{\msb}~ + ~C_{2,q}^{\msb} \otimes q^{\msb}~ + ~C_{2,g}^{\msb} \otimes g^{\msb},
\label{eq:q}
\eeq
while to obtain the gluon we take
\beq
g^{\rm DIS}~~=~~g^{\msb}~ - ~C_{2,q}^{\msb} \otimes q^{\msb}~ - ~C_{2,g}^{\msb} \otimes g^{\msb}.
\label{eq:g}
\eeq
The last transformation is not unique.  However it represents the simplest and most natural
choice to maintain the 100\% momentum carried by the partons.
Indeed, this is the conventional choice which has been used in the past to 
obtain DIS-scheme parton distributions, see for example Refs. \cite{DISdefa,GRV94,MRSTDIS,CTEQ6}. 

At high $x$, the term $C^{\msb}_{2,g} \otimes g^{\msb}$ is effectively
negligible.  The coefficient function $C^{\msb}_{2,q}$ must be
consistent with the Adler sum rule, and hence it has a vanishing zeroth
moment (consistent with quark number conservation).  However the perturbative coefficients
give a large positive contribution at high $x$, behaving as [ln$^{2n-1}(1-x)/(1-x)]_+$ at
order $\alpha^n_s$.  Hence the term $C^{\msb}_{2,q} \otimes q^{\msb}$ plays a crucial
role at high $x$.

Although the partons are significantly different in the two schemes, the jet cross section is
rendered unchanged up to NLO by a compensating change in the hard subprocess cross sections.
To see this we note that the total jet cross section may be written schematically
as\footnote{For simplicity, it is sufficient in this discussion to ignore the difference
between quarks and antiquarks.}
\beq
\sigma_{\rm jet}~=~\sigma^i_{qq} \otimes q^i \otimes q^i~+~\sigma^i_{qg} \otimes q^i \otimes g^i~+~
\sigma^i_{gg} \otimes g^i \otimes g^i
\eeq
with $i=\msb$ or DIS.   Thus, using (\ref{eq:q}) and (\ref{eq:g}) with the final term neglected, we
find, up to NLO, that
\begin{eqnarray}
\sigma^{\dis}_{qq} & = & \sigma^{\msb}_{qq}-2\sigma^{\msb}_{qq}\otimes C^{\msb}_{2,q}
+\sigma^{\msb}_{qg}\otimes C^{\msb}_{2,q}\\
\sigma^{\dis}_{qg} & = & \sigma^{\msb}_{qg}+2\sigma^{\msb}_{gg}\otimes C^{\msb}_{2,q}
-\sigma^{\msb}_{qg}\otimes C^{\msb}_{2,q}\\
\sigma^{\dis}_{gg} & = & \sigma^{\msb}_{gg}.
\end{eqnarray}
As a result the increase in the high $x$ quark density is compensated by a decrease in the
hard subprocess cross section, and the quark-dependent decrease in the gluon is compensated by an 
increase in the quark-gluon cross-section.

We can now explain the improvement in the quality of the description of the jet data using
the DIS scheme that was noted by Klasen and Kramer \cite{KK}.   They used the CTEQ3M($\msb$)
and CTEQ3D(DIS) partons in their analysis.  The difference between these partons can be seen
in Fig. 2 of Ref.~\cite{KK}.  These partons were determined by CTEQ in separate global fits
performed in the
two schemes.  The precise structure function data at high $x$ forces the quarks to satisfy
(\ref{eq:q}) to good accuracy.  On the other hand, at the time of these CTEQ fits \cite{CTEQ3} there was
no strong constraint on the high $x$ gluon, and consequently it is very similar in the two schemes,
clearly in contradiction with (\ref{eq:g}) (and with the CTEQ6 results shown
in Fig.~\ref{fig:c6DM} above).  Hence the
increased hard subprocess cross section $\sigma^{\dis}_{qg}$ was not accompanied by a
decrease in the gluon distribution, and the prediction for the high $E_T$ jet cross section
increased significantly.  However the more precise data that are available now forces the
gluon to, at least approximately, respect the transformation relation given in (\ref{eq:g}).
Nevertheless, the complicated nature of the transformation may result in differences in the
fits to the data in the two schemes due to the 
simplicity of the form of the gluon parameterization
at high $x$.

The DIS factorization scheme is certainly more natural for quarks.  The $\msb$ scheme was
devised to be particularly simple when using the standard, but unphysical,
dimensional regularization procedure for regularization of infrared singularities.  
Moreover if, as expected, the
high $x$ valence quarks dominate the high $x$ gluon in the DIS scheme\footnote{Recall that if
this dominance occurred in the $\msb$ scheme, then the high $x$ gluon is negative
in the DIS scheme.}, then, according
to transformation (\ref{eq:g}), the $\msb$ gluon in the high $x$ limit is determined by
the behaviour of the valence quarks
\beq
g^{\msb}~~\simeq~~g^{\rm DIS}~ + ~C_{2,q}^{\msb} \otimes q^{\msb}.
\label{eq:gms}
\eeq
It is therefore natural to adopt the following procedure. We parameterize the DIS gluon
at the input scale so that its large $x$ behaviour is governed by the conventional form 
$(1-x)^{\eta_g({\rm DIS})}$.
Then, as usual, we perform the global fit in the $\msb$ scheme, but now with the input gluon
parameterized according to (\ref{eq:gms}).  To be precise we take
\beq
g^{\msb}(x,Q^2_0)~~=~~g^{\rm DIS}(x,Q^2_0)~ + ~C_{2,{\rm NS}}^{\msb} \otimes \sum_{q=u,d} q_{\rm val}^{\msb}(x,Q^2_0),
\label{eq:ginput}
\eeq
with $Q^2_0=1 ~\GeV^2$. We note that our input gluon has exactly the same number of parameters as
usual. At NLO the non-singlet coefficient function is
\begin{eqnarray} \label{eq:coeff}
C_{2,{\rm NS}}^{\overline{\mbox{\scriptsize MS}}}(x) & = &
\frac{\alpha_s C_F}{2\pi}\left[ 2\left( {\ln(1-x)\over 1-x} \right)_+
- \frac{3}{2}\left( {1\over 1-x} \right)_+ -(1+x)\ln(1-x) \right.
\nonumber \\
& & \left.
 -{ 1+x^2 \over 1-x} \ln{x}+ 3 + 2x -\left(
\frac{\pi^2}{3}+\frac{9}{2}\right)
\delta(1-x) \right]. \
\end{eqnarray}
Thus, for example, if $q_{\rm val}^{\msb}$ goes like $A(1-x)^n$ at high $x$, then the
convolution in (\ref{eq:ginput}) gives a behaviour
\beq
g^{\msb}~~\sim~~\frac{\alpha_s C_F}{2\pi}~\ln^2(1-x)~A(1-x)^n
\eeq
for the `valence-driven' gluon at high $x$.  That is a ${\rm log}^2$ enhancement over the
fall-off of the valence quark.
The NNLO expression of the coefficient function can be found in Ref.~\cite{ZN},
and leads to a leading-log $\ln^4(1-x)$ enhancement. In principle a large-$x$
resummation \cite{softglu} could be performed.  
We shall see that the structure of the input form, (\ref{eq:ginput}),
of the gluon cannot easily be mimicked by a direct $\msb$ gluon parameterization.  It turns out to
be important that the high $x$ gluon is driven by the valence quarks.

We perform global analyses at both NLO and NNLO using the standard cuts on the data $(Q^2>2~\GeV^2$ and
$W^2>12.5~\GeV^2$).  We use the parameterization of Ref.~\cite{MRST2001}, except that the gluon is
first parametrised in the DIS scheme and then transformed according to (\ref{eq:ginput}).
Indeed, the NLO global analysis with this new gluon parameterization works extremely well, and
is even better for the NNLO DGLAP fit. When we performed our previous NNLO analyses \cite{MRSTnnlo,
MRSTerror1,MRSTerror2} the complete set of splitting functions was not available, at this order,
and we used the bounds on their behaviour obtained by van Neerven and Vogt \cite{NV}.  However
in the present NNLO analysis we use the splitting functions which have recently become
available \cite{MVVns,MVVs}. Since these exact functions lie approximately centrally
within the original bounds, the NNLO partons are essentially unaltered.

\begin{figure}
\mbox{\epsfig{figure=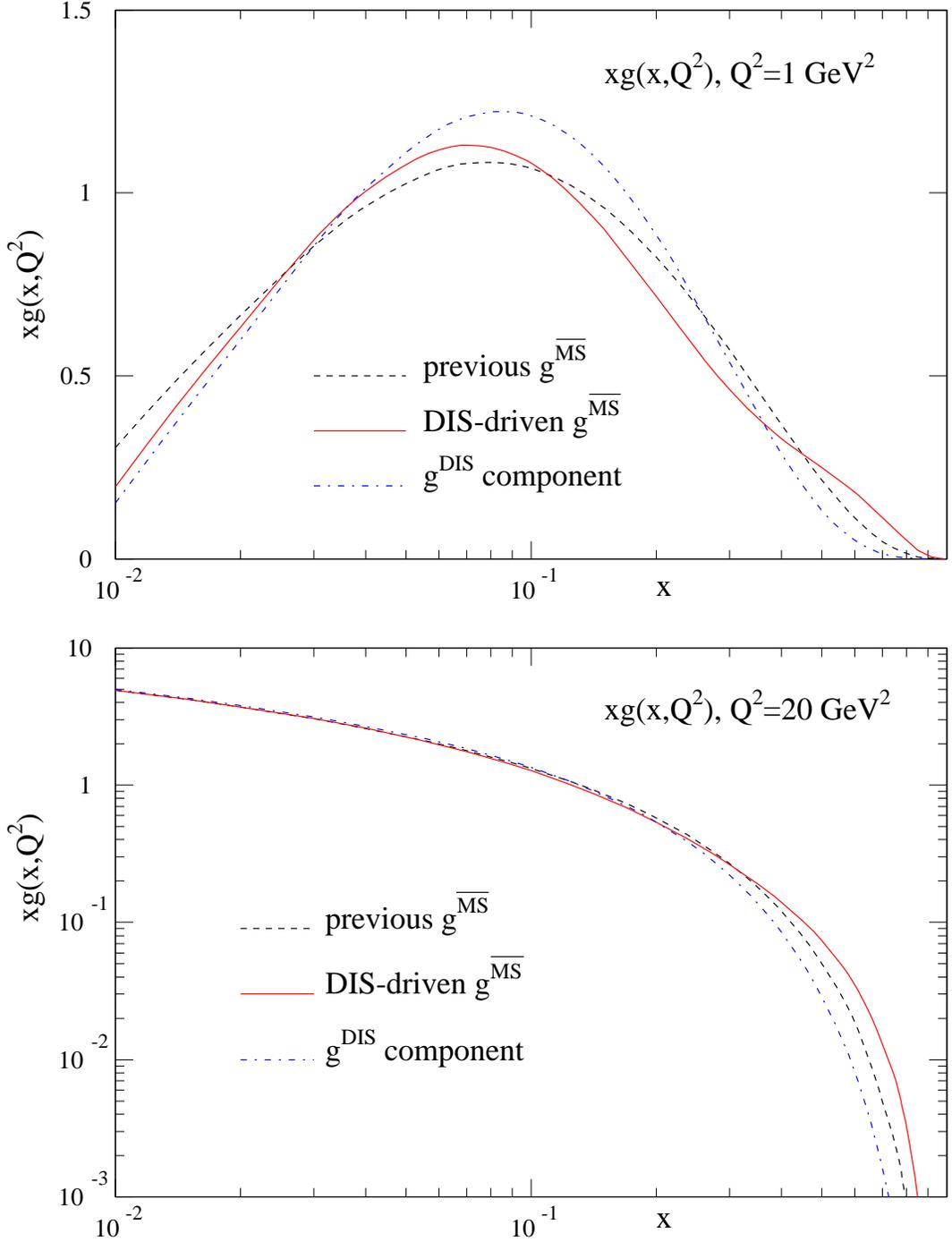,width=15cm}}
\vspace{-2.5cm}
\caption{The previous (default) MRST NLO $\msb$-scheme
gluon compared to
that obtained when the high $x$ behaviour of the gluon is determined by the quark
transformation between $\msb$ and DIS schemes, as in (\ref{eq:gms}).  The two
gluons are shown, respectively, by continuous and dashed curves.  Also shown
by dot-dashed curves is this latter
gluon when transformed to the DIS scheme.  Exactly the same data sets are used
in the two fits.}
\label{fig:glujet}
\end{figure}

First, consider the NLO analysis.  Our most recent
default gluon\footnote{Since the global analysis of Ref.~\cite{MRSTerror1} was performed,
we now include in the fit the new NuSea data for Drell-Yan production
in $pp$ collisions \cite{E866}, the high-$Q^2$ 1999-2000
ZEUS data for $F_2$ \cite{ZEUSnc} and the charged-current HERA data \cite{HERAcc}.
This leads to only minor changes in the partons, but the gluon parameter
$\eta_g (\msb)$ decreases slightly
from 3.15 to 2.98, and $\alpha_s(M_Z^2)$ increases slightly to 0.1200.
  However the new Drell-Yan data on a proton target turn out to be
more compatible with the Tevatron jet data than the previous Drell-Yan nuclear target
data.}
 behaves like $(1-x)^{2.98}$, that is $\eta_g (\msb) = 2.98$,
corresponding to a $\chi ^2 =154$ description of the D0 and CDF inclusive jet $E_T$ distributions.
If, now, we perform a NLO fit with the ($\msb$) gluon parameterized according to (\ref{eq:gms}) then
the description of the jet data is considerably improved, with $\chi ^2 =116$, while
$\chi ^2$ for the remainder of the data only increases by 12.
Interestingly, with the new parameterization
the $g^{\rm DIS}$ component in (\ref{eq:gms}) behaves as $(1-x)^{4.5}$, much more consistent
with the simple counting rule expectations, (\ref{eq:count}).  The resulting `DIS-driven' $\msb$ gluon is
compared to our previous default $\msb$ gluon in Fig.~\ref{fig:glujet} at $Q^2 = 1$ and
$Q^2 = 20 ~{\rm GeV}^2$.
The two gluons are shown by continuous and dashed curves respectively.  We see that the DIS-driven
gluon is considerably larger at very high $x$ (due its quark component), and has a
shoulder-like structure at the input scale.  The dot-dashed curves show the
form of the $g^{\rm DIS}$ component of (\ref{eq:gms}), which clearly has a more natural
$(1-x)$ behaviour than our previous default gluon.  In this new NLO analyses the
value of $\alpha_s(M_Z^2)$ has increased slightly from 0.1200 to 0.1205, since the increase of the gluon at
very high $x$ results in a decrease for $x \sim 0.1$, and so the coupling has to increase
to fit the NMC and HERA $F_2$ data.
\begin{figure}
\mbox{\hskip -1cm\epsfig{figure=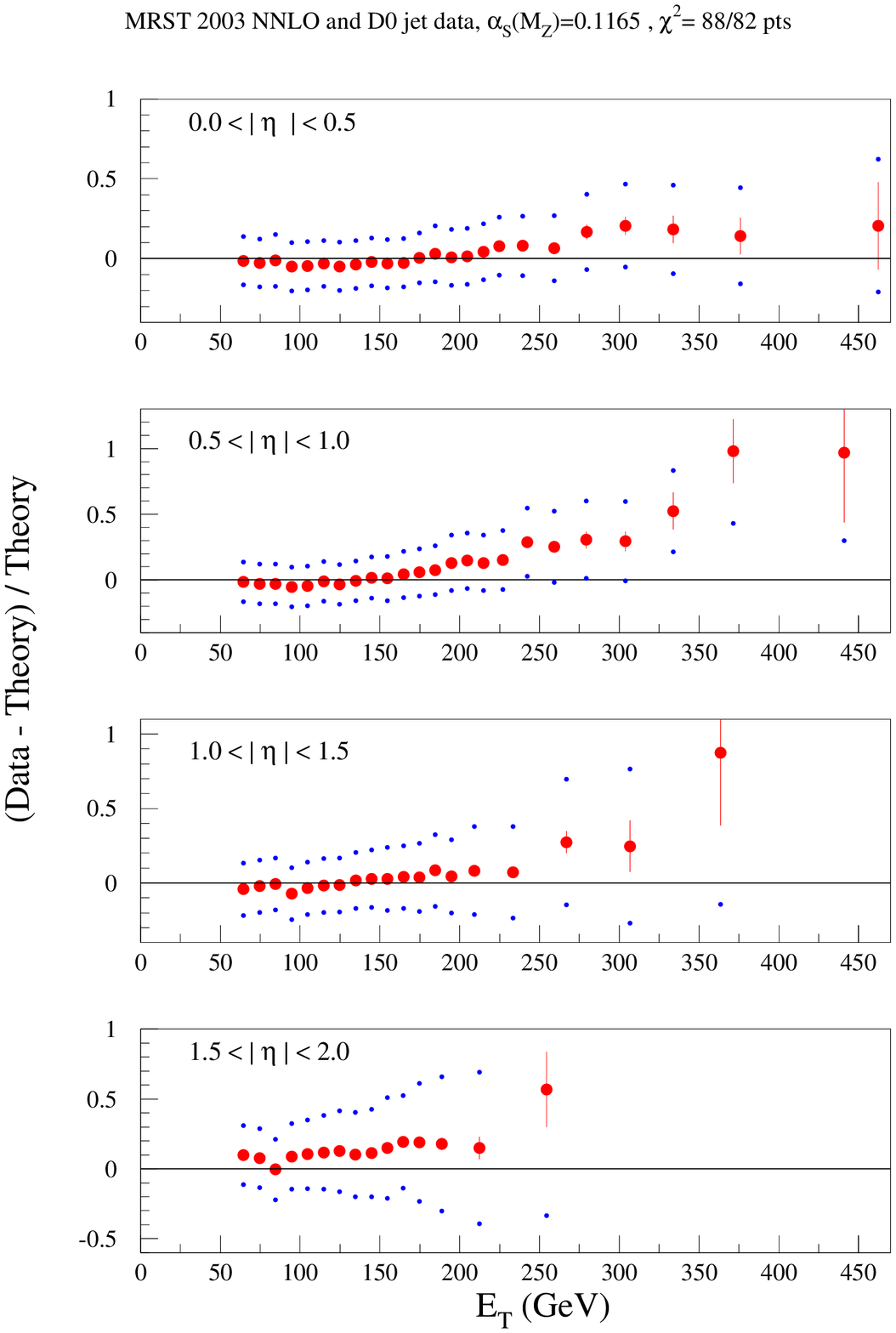,width=8.6cm}\epsfig{figure=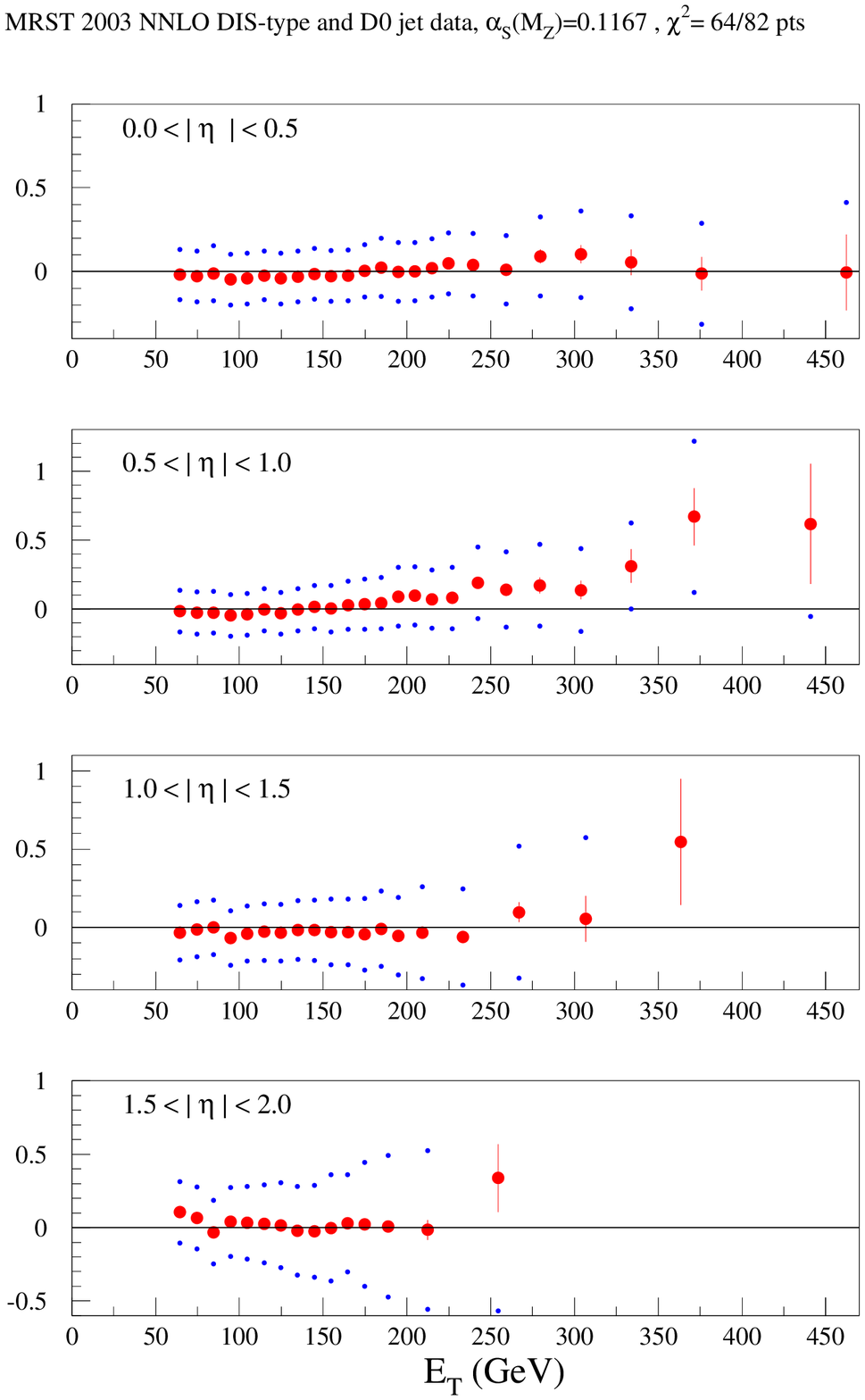,
width=8.6cm}}
\vspace{-1.3cm}
\caption{The description of the D0 inclusive jet $E_T$ distributions
in different rapidity intervals \cite{D0} obtained in our standard default NNLO analysis compared
to the improvement obtained using the new gluon parameterization of (\ref{eq:ginput}).
The bands indicate the allowed shifts from the central value for each data point
obtained by adding the correlated errors in quadrature.  The `valence-quark driven'
parametrization of the gluon improves $\chi^2$ for the description of the D0 data
from 88 to 64.}
\label{fig:jet}
\end{figure}
\begin{figure}
\mbox{\epsfig{figure=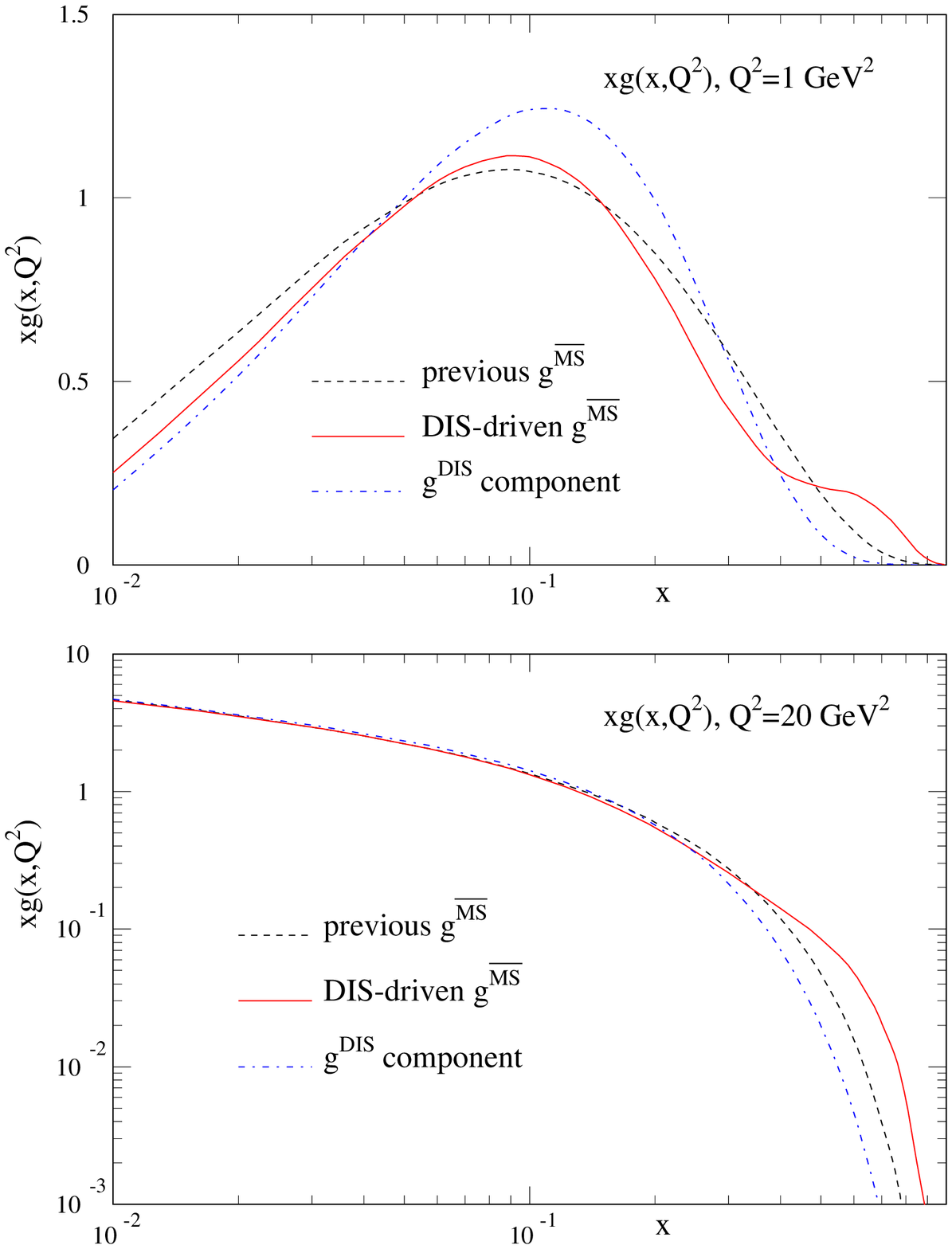,width=15cm}}
\vspace{-2.5cm}
\caption{The previous default MRST NNLO $\msb$-scheme
gluon compared to
that obtained when the high $x$ behaviour is determined by the quark
transformation between $\msb$ and DIS schemes. Also shown is this latter
gluon when transformed to the DIS scheme.}
\label{fig:glujetnnlo}
\end{figure}

The improvement in the NNLO global fit is
even better than that at NLO, when the DIS-driven gluon parameterization
is used.  Now, $\chi ^2$ for the description of the D0 and CDF
inclusive jet $E_T$ distributions is reduced from 164 to 117, with the overall $\chi ^2$
of the global fit
decreasing by 79.  We illustrate the improvement in Fig.~\ref{fig:jet}
by comparing the default and the new fits to the
inclusive jet $E_T$ distributions measured by the D0 Collaboration \cite{D0}.  The improvement in
the description of the CDF inclusive jet data \cite{CDF} is similar.
At NNLO, not only the fit to the jet data, but also to the HERA data, is
improved by the new parameterization; or more precisely the relaxation of the tension between the two
data sets allows the description of both to improve at NNLO\footnote{The analysis is
repeated with various cuts on $x$ and $Q^2$ to see whether the improvement in fit quality after cuts 
have been applied is reduced by the introduction
of the new parameterization. At NLO, when conservative cuts \cite{MRSTerror2} of $x=0.005$ and
$Q^2=10~\GeV^2$ are applied and a new fit performed, for the standard parameterization 
the refit results in an improvement in $\chi^2$ of
79 compared to the partons obtained from the fit with the default cuts ($x=0$ and
$Q^2=2~\GeV^2$). When this procedure is repeated with the new parameterization for the high-$x$ gluon  
the improvement due to the refit is reduced to 54. At NNLO, with conservative cuts of $x=0.005$ and 
$Q^2=7~\GeV^2$ the refitting procedure with the standard parameterization gives an improvement in 
$\chi^2$ of 79, and this is reduced to an improvenent with refitting of 41 when the new 
parameterization is used. Hence, in neither case can the new parameterization be said to 
remove the improvement with refitting after cuts are applied. Nevertheless, the reduction in $\chi^2$ 
with refitting comes about in essence due to more gluon moving to high
$x$ when it is allowed to, and the improvement in the shape of the high-$x$ gluon in the new 
parameterization clearly moderates this effect.}.
Also, in this case there is even less change in $\alpha_S(M_Z^2)$
when introducing the new parameterization; it increases from 0.1165 to 0.1167.
The new NNLO  gluon is compared to our previous NNLO gluon in Fig.~\ref{fig:glujetnnlo}.
The shoulder at high $x$ is even more pronounced; the additional quark contribution,
$C_{2,q}^{\msb ,(2)} \otimes q^{\msb}$ is positive and significant at very high $x$, so
the high $x$ NNLO gluon is even more determined by the quark distributions than that at NLO.

To conclude, there is an inherent instability in the size and shape of the gluon at high $x$
-- it changes dramatically as one goes from one factorization scheme to another.
The natural assumption that the high-$x$ gluon should be smooth, with the
usual $(1-x)^{\eta_g}$ behaviour at high $x$, in the DIS scheme, results in a
relatively large high-$x$ gluon with structure in
the $\msb$ scheme.  This is exactly what is needed to give an excellent
description of the Tevatron jet data.   Indeed, using the quark-driven
gluon parametrization given by (\ref{eq:ginput}), we find a much improved fit to jet data at NLO, and
a dramatic improvement in the fit to both the jet data and the total global fit at NNLO where
the scheme dependence increases still further.
The main reason for the improvement can be traced to the discussion of the
description of the Tevatron jet data in Ref.~\cite{MRSTnnlo}.  From the viewpoint of the DIS factorization
scheme, the good fit to the jet data is driven by large valence quarks at high $x$, and
a naturally smaller and smooth gluon.  In fact it was already noticed that
in a LO fit, where the quarks are very similar to those in the DIS scheme, a good description of the
Tevatron jet data could be obtained ($\chi^2=123$), with an input
gluon behaving as $(1-x)^{6.49}$ at high $x$ \cite{MRSTnnlo}.
Thus, it is a pleasing, and seemingly natural outcome that the best NLO and NNLO 
fits\footnote{These parton sets,
which we denote by MRST2004, can be
found at \hfil\break http://durpdg.dur.ac.uk/hepdata/mrs.html} (performed in the $\msb$ scheme) come
from a high-$x$ gluon of the form we would intuitively expect in the more physically motivated
DIS factorization scheme. However, even if one does not believe that there is any reason for the DIS-scheme
gluon to be the more physical at high $x$, the procedure in this paper provides an extremely
successful way 
to obtain a high-$x$ gluon of precisely the size and shape needed by the Tevatron jet data within a 
global fit.

\section*{Acknowledgements}

 RST would like to thank
the Royal Society for the award of a University Research Fellowship. ADM and RGR
would both like to thank the Leverhulme Trust for the award of an Emeritus
Fellowship. The IPPP gratefully acknowledges financial support from the UK
Particle Physics and Astronomy Research Council.\\

\vspace{-0.6cm}

%\newpage

\end{document}